\documentclass[twocolumn,english,showpacs,showkeys,superscriptaddress,citeautoscript]{revtex4-2}
\usepackage[T1]{fontenc}
\usepackage[latin9]{inputenc}
\usepackage[active]{srcltx}
\usepackage{amsmath}
\usepackage{amssymb}
\usepackage{graphicx}
\usepackage{wasysym}
\usepackage{esint}

\makeatletter

\providecommand{\tabularnewline}{\\}

\usepackage{babel}
\usepackage{rotating}

\makeatother

\usepackage{babel}
\begin{document}
\title{Emergence of quantum spin frustration in spin-1/2 Ising-Heisenberg
model on a decorated honeycomb lattice}
\author{Onofre Rojas}
\affiliation{Department of Physics, Institute of Natural Science, Federal University
of Lavras, 37200-900, Lavras -MG, Brazil}
\begin{abstract}
We study the spin-1/2 Ising-XXZ model on a decorated honeycomb lattice
composed of five spins per unit cell, one Ising spin, and four Heisenberg
spins. This model involving the Heisenberg exchange interaction is
one of the few models that can be exactly solvable through the generalized
star-triangle transformation. The significance of this model is its
close relationship to the fully decorated quantum Heisenberg honeycomb
lattice since 4/5 of the particles are Heisenberg spins. We investigate
the phase diagram at zero temperature and identify a relevant quantum
spin frustrated phase resulting from the contribution of quantum Heisenberg
exchange interaction. We obtain an exact residual entropy for the
quantum spin frustrated phase, which coincides with the residual entropy
of the antiferromagnetic spin-1/2 Ising model on a triangular lattice.
We also thoroughly explore its thermodynamic properties, focusing
mainly on the frustrated region such as entropy, specific heat, spontaneous
magnetization, and critical temperature under several conditions.
\end{abstract}
\maketitle

\section{Introduction}

One of the topics of intense research in statistical physics is frustrated
spin systems, showing numerous fascinating phenomena. For example,
the role of frustration in the emergence of magnetic phases and the
nature of phase transitions received a great deal of attention in
the last decades. In this context, lattices of quantum spin models
are relevant, like the Heisenberg model that describes in detail several
synthesized compounds. Typically, spin models in statistical physics
cannot be solved exactly, so most of them can only be studied numerically.
Therefore, the exact solutions were only achieved in limited cases.
However, after the solution found by Onsager for the two-dimensional
(2D) Ising model\citep{onsager}, inspired several attempts to solve
other similar models. Such as the honeycomb lattice\cite{Wu-jmp73,Horiguchi-Wu,Kolesik},
whose exact solution of a honeycomb lattice with an external magnetic
field was provided by Wu\citep{wu-90}. Besides, kagomé lattice was
also widely discussed in the literature \citep{azaria,lu-wu} and
reference therein.

One of the candidates for obtaining spintronic devices with high performance
and high integration density may be 2D semimetallic materials, thus
becoming highly desirable. Based on first-principles calculations,
was predicted a potential candidate of 2D half-metals, named monolayer
$\mathrm{Mg}_{3}\mathrm{Si}_{2}$ \citep{song}, which has a honeycomb-kagomé
lattice pattern. Also proposed was a 2D-silicon semiconductor by introducing
kagomé topology into a honeycomb lattice, i.e., a hybrid honeycomb-kagomé(hhk)
structure\citep{sang}, which demonstrates high geometric stability
and excellent semiconducting properties of the hhk-silicene. Other
theoretical investigations on honeycomb-kagomé lattice appear in a
variety of compounds, such as 2D honeycomb-kagomé $\mathrm{Be}_{3}\mathrm{Pb}_{2}$
\citep{zhang}. This compound exhibits a nontrivial topology in the
electronic structure that accompanies the spin-orbit coupling (SOC)
induced by the energy gap bandwidth. Recently, several other investigations
have been developed, with a number of exotic properties\citep{mizoguchi,L-zhang-18,L-zhang-19,Y-ding}.

Normally, geometric frustration occurs within a triangular spin structures.
When the antiferromagnetic exchange interactions cannot be satisfied
simultaneously, we achieve a considerable degenerated ground state
that leads to a frustration. In such a way, frustrated magnets have
been attracted great scientific interest because of quantum spin liquid
in 2D systems, which has been pointed to play a striking role in high-temperature
superconductors. In this sense, there are several investigations concerning
geometrically frustrated kagomé lattice compounds, which are challenging
topics. Some materials with spin-1/2 kagomé structures present the
liquid state of quantum spin at zero temperature, such as the antiferromagnetic
kagomé lattice. Like atacamite family\citep{shores} whose general
formula is $\mathrm{Zn}_{x}\mathrm{Cu}_{4-x}\mathrm{(OH)}_{6}\mathrm{Cl}_{2}$
with $(0\leqslant x\leqslant1)$, which follows a regular kagomé lattice
structure. Using the single-crystal samples of the antiferromagnet
spin-1/2 kagomé-lattice\citep{Han} $\mathrm{ZnCu}_{3}\mathrm{(OD)}_{6}\mathrm{Cl}_{2}$
(also called herbertsmithite), certain relevant features of the quantum
spin liquid phase were observed.

Several decorated spin models can be solved by applying the well-known
decoration transformation introduced in the 1950s by M. E. Fisher\citep{Fisher}
and Syozi\citep{syozi}. Later generalized for arbitrary spins, such
as the classical or quantum spin models\citep{PhyscA-09,strecka-pla,Roj-sou11}.
This transformation is essential because we can map cumbersome models
into simple or exactly solvable models. In what follows, we mention
a few typical examples of this approach: the geometrical frustrated
Cairo pentagonal lattice Ising model\citep{M-cairo}, the geometrically
frustrated Ising-XXZ spin on expanded kagomé lattice\citep{kg-rojas},
the XXZ-Ising model on the triangular kagomé lattice with spin-1/2\citep{loh,strecka-triang},
and mixed-spin Ising model on a honeycomb lattice\citep{torrico-triang}.
Furthermore, 2D square-hexagon (denoted for simplicity by 4-6) XXZ-Ising
with the spin-1/2 model was investigated using the same approach\citep{our-4-6-latt}.
Similarly, two-dimensional 3-12 lattice is also known in the literature\citep{Lin-3-12,barry95}
as the star lattice, Fisher lattice, expanded kagomé lattice, or even
triangular honeycomb lattice. The study of these models was based
on their close relationship with geometrically frustrated magnetic
polymers\citep{octa-kagome}. There are even several investigations
concerning other Ising-Heisenberg lattices like bipyramidal plaquette\cite{galisova21,Galisova},
triangles-in-triangles lattices\cite{cisarova-13,Michaud}, martini
lattice\cite{zad}, and triangulated Husimi\cite{Streck-15} and kagomé\cite{Strecka-20}
lattices. It is also worth mentioning also that there are investigations
on spin-1/2 Heisenberg antiferromagnetic star lattice structure\cite{Richter04,Yan-10}.

The outline of this report is as follows. In Sec.2, we present the
spin-1/2 Ising-XXZ model on a decorated honeycomb lattice, so we map
through the generalized star-triangle transformation technique into
an effective triangular lattice. And subsequently, we investigate
the ground-state phase diagram, showing an unconventional quantum
spin frustration. Whereas, in Sec.3, we report the thermodynamics
result for studying entropy, specific heat, critical temperature,
and spontaneous magnetization. Finally, Sec.4 focuses on presenting
our concussions.

\section{Ising-XXZ model on a decorated honeycomb lattice}

\begin{figure}[h]
\includegraphics[scale=0.8]{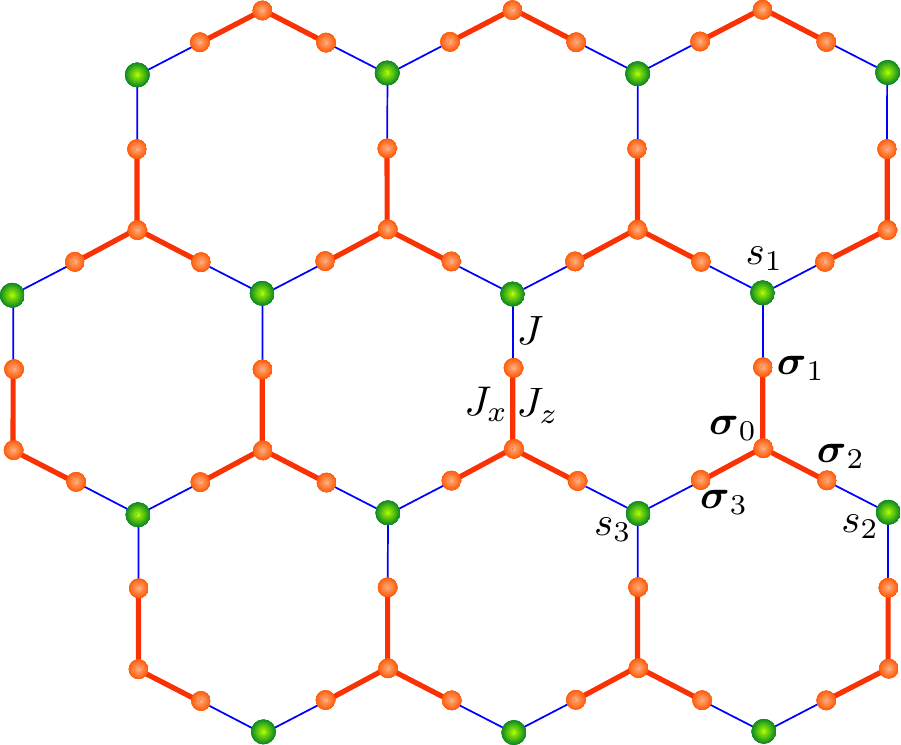}

\caption{\label{fig:Honeycomb-lattice}Spin-1/2 Ising-Heisenberg decorated
honeycomb lattice, orange spheres correspond to Heisenberg spins,
and green spheres denote Ising spins. Orange lines correspond to Heisenberg
exchange interactions, and blue lines indicate the Ising exchange
interactions.}
\end{figure}

A classical version of the spin-1/2 Ising model on a decorated honeycomb
lattice has been considered by Syozi\cite{syozi} and later Wu\cite{wu-90}
investigated in the presence of an external magnetic field. However,
before considering its fully quantum version, a natural generalization
would be the spin-1/2 Ising-XXZ model on a decorated honeycomb lattice,
as illustrated in Fig.\ref{fig:Honeycomb-lattice}, by Heisenberg
spin (small sphere) and Ising spin (large sphere). Thus, the thick
solid line between small spheres represents the exchange interactions
of Heisenberg spin, whereas the thin blue solid line bonds small and
large spheres corresponding to Ising-type exchange interactions.

The Hamiltonian that describes the decorated honeycomb lattice (see
Fig.\ref{fig:Honeycomb-lattice}), can be expressed as follows

\begin{equation}
H=-\negthickspace\sum_{<i,j>}\hspace{-0.2cm}\left[J_{x}\negmedspace\left(\sigma_{i}^{x}\sigma_{j}^{x}\!+\!\sigma_{i}^{y}\sigma_{j}^{y}\right)\!+\!J_{z}\sigma_{i}^{z}\sigma_{j}^{z}\right]-J\hspace{-0.2cm}\sum_{<i',l>}\negthickspace\sigma_{i'}^{z}s_{l},\label{eq:Ham-1}
\end{equation}
where, by $\sigma^{\alpha}$ we denotes the Heisenberg spin operator,
with $\alpha=\{x,y,z\}$, and $s_{l}$ denotes the Ising spin $s_{l}=\pm1/2$.
The first summation corresponds to the anisotropic Heisenberg exchange
interaction between the nearest neighbor spins, with $J_{x}$ standing
to the Heisenberg exchange interaction in $xy$-component, and $J_{z}$
corresponds to Heisenberg exchange interaction in the $z$-component.
The second summation indicates the Ising spin exchange interaction
$J$ between nearest Ising and Heisenberg spins.

\subsection{Generalized star-triangle transformation}

\begin{figure}[h]
\includegraphics[scale=0.8]{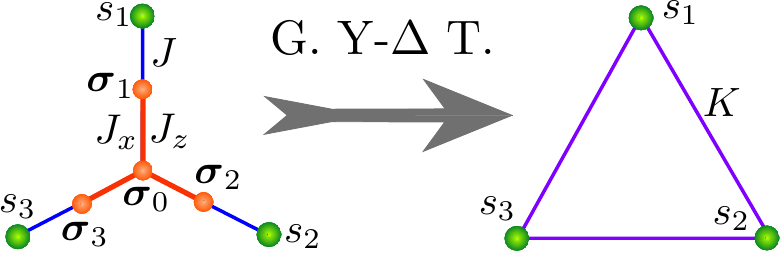}

\caption{\label{fig:3-leg-hybrid-star}Schematic representation of generalized
star-triangle transformation (G. Y-$\Delta$ T.) of a spin-1/2 Ising-XXZ
model of decorated three-leg star onto a spin-1/2 Ising model on a
triangular structure.}
\end{figure}

The model proposed in \eqref{eq:Ham-1} could be viewed as a decorated
Ising-XXZ model that can be mapped exactly by means of a generalized
generalized star-triangle transformation\citep{Fisher,syozi,PhyscA-09,strecka-pla,Roj-sou11},
as illustrated in Fig.\ref{fig:3-leg-hybrid-star}, which basically
makes a partial trace over all Heisenberg spins, leaving only the
Ising spins interaction between the nearest neighbors. Thus, we will
consider a three-legged hybrid star system as schematically represented
in Fig.\ref{fig:3-leg-hybrid-star}(left). The three internal bonds
of the star are described by $J_{x}$ and $J_{z}$ corresponding to
Heisenberg-type interactions, while the external legs are Ising-type
interactions $J$, as shown in Fig.\ref{fig:3-leg-hybrid-star}(left).
Hence, the Ising-XXZ model on decorated honeycomb lattice can be mapped
onto a triangular lattice with effective Ising spin coupling $K$
through a generalized star-triangle transformation approach \citep{Fisher,syozi,PhyscA-09,strecka-pla,Roj-sou11},
as displayed in Fig.\ref{fig:3-leg-hybrid-star}(right).

To carry out the lattice transformation, let us define the following
operator 
\begin{equation}
\mathbf{V}(\{\boldsymbol{\sigma},s\})=\mathrm{e}^{\beta\overset{3}{\underset{i=1}{\sum}}\left[J_{x}\left(\sigma_{i}^{x}\sigma_{0}^{x}+\sigma_{i}^{y}\sigma_{0}^{y}\right)+J_{z}\sigma_{i}^{z}\sigma_{0}^{z}+J\sigma_{i}^{z}s_{i}\right]},
\end{equation}
where $\{\boldsymbol{\sigma},s\}$ denote $\{\boldsymbol{\sigma}_{0},\boldsymbol{\sigma}_{1},\boldsymbol{\sigma}_{2},\boldsymbol{\sigma}_{3},s_{1},s_{2},s_{3}\}$,
and $\boldsymbol{\sigma}=\{\sigma^{x},\sigma^{y},\sigma^{z}\}$. The
spin index $s_{l}$ in Hamiltonian \eqref{eq:Ham-1}, was reindexed
by $s_{i}$ (same index of $\sigma_{i}$) just for simplicity, without
imposing any restrictions. Thus, the Boltzmann factor of the decorated
hybrid-spin model becomes 
\begin{equation}
w(\{s\})=\mathrm{tr}_{\{\sigma\}}\left[\mathbf{V}(\{\boldsymbol{\sigma},s\})\right].
\end{equation}

The inner star system (decorated) is expressed as Heisenberg coupling,
and it provides two configurations for Ising spins (legs); these correspond
to the following configurations $\{+++\}$ and $\{++-\}$ (since the
lattice is invariant under spin inversion). So by defining $\varsigma=s_{1}+s_{2}+s_{3}$,
the pair configurations become $\varsigma=3/2$ and $1/2$. A convenient
way to obtain the trace over Heisenberg spin is diagonalizing the
$16\times16$ matrix for both Ising spin configurations $\varsigma=3/2$
and $\varsigma=1/2$. In Table \ref{tab:Heisenberg-spin-eigenvalues,}
is reported the 16 eigenvalues for both configurations. Some eigenvalues
are large expressions, so we have denoted conveniently by using the
following notations: 
\begin{alignat}{1}
A_{\pm}= & \left(J\pm2J_{z}\right)^{2}+12J_{x}^{2},\label{eq:A}\\
B_{\pm}= & 5\left(2J_{x}^{2}+J^{2}\right)\pm2\sqrt{\left(2J^{2}+3J_{x}^{2}\right)^{2}+16J^{2}J_{x}^{2}},\label{eq:B}\\
Q= & \tfrac{1}{12}J^{2}+\tfrac{1}{9}J_{z}^{2}+\tfrac{1}{4}J_{x}^{2},\label{eq:Q}\\
R_{\pm}= & \left(\tfrac{1}{8}J_{z}\mp\tfrac{1}{16}J\right)J_{x}^{2}+\tfrac{1}{27}J_{z}^{3}-\tfrac{1}{12}J^{2}J_{z},\label{eq:R}\\
\theta_{\pm}= & \arccos\left(\tfrac{R_{\pm}}{\sqrt{Q^{3}}}\right).\label{eq:theta}
\end{alignat}
Indeed, the corresponding eigenvectors are also relevant, but we do
not express them explicitly because they involve huge cumbersome algebraic
expressions. Nevertheless, further on, we will use some of these eigenvectors
mainly to study the quantum spin frustrated phase, and the thermal
magnetization of the model.

Now we write the Boltzmann factor for the case $\varsigma=3/2$, so
for simplicity, we denote by $w(3/2)=w_{3}$,
\begin{equation}
w_{3}=\sum_{k=1}^{16}{\rm e}^{-\beta\varepsilon_{k}^{(3)}},\label{eq:w(3/2)}
\end{equation}
where $\varepsilon_{k}^{(3)}$ means the eigenvalues of Heisenberg
spins, which are given in table \ref{tab:Heisenberg-spin-eigenvalues,}
(second column).

Similarly, for $\varsigma=1/2$, the Boltzmann factor becomes
\begin{equation}
w_{1}=\sum_{k=1}^{16}{\rm e}^{-\beta\varepsilon_{k}^{(1)}},\label{eq:w(1/2)}
\end{equation}
and its eigenvalues $\varepsilon_{k}^{(1)}$ are given in Table \ref{tab:Heisenberg-spin-eigenvalues,}
(third column).

\begin{table}
\begin{tabular}{|c|c|c|}
\hline 
$k$  & $\varepsilon_{k}^{(3)}$ for $\varsigma=3/2$  & $\varepsilon_{k}^{(1)}$ for$\varsigma=1/2$\tabularnewline
\hline 
\hline 
1  & $-\frac{3}{4}\left(J_{z}+J\right)$  & $-\frac{1}{4}\left(3J_{z}+J\right)$\tabularnewline
\hline 
2  & $-\frac{3}{4}\left(J_{z}-J\right)$  & $-\frac{1}{4}\left(3J_{z}-J\right)$\tabularnewline
\hline 
3  & $-\frac{1}{4}\left(J_{z}+J\right)$  & $-\frac{1}{4}\left(J_{z}+J\right)$\tabularnewline
\hline 
4  & $-\frac{1}{4}\left(J_{z}+J\right)$  & $-\frac{1}{4}\left(J_{z}+J\right)$\tabularnewline
\hline 
5  & $-\frac{1}{4}\left(J_{z}-J\right)$  & $\frac{1}{4}\left(J_{z}+\sqrt{J^{2}+4J_{x}^{2}}\right)$\tabularnewline
\hline 
6  & $-\frac{1}{4}\left(J_{z}-J\right)$  & $\frac{1}{4}\left(J_{z}-\sqrt{J^{2}+4J_{x}^{2}}\right)$\tabularnewline
\hline 
7  & $\frac{1}{4}\left(J_{z}+\sqrt{J^{2}+4J_{x}^{2}}\right)$  & $\frac{1}{4}\left(J_{z}+\sqrt{B_{+}}\right)$\tabularnewline
\hline 
8  & $\frac{1}{4}\left(J_{z}+\sqrt{J^{2}+4J_{x}^{2}}\right)$  & $\frac{1}{4}\left(J_{z}-\sqrt{B_{+}}\right)$\tabularnewline
\hline 
9  & $\frac{1}{4}\left(J_{z}-\sqrt{J^{2}+4J_{x}^{2}}\right)$  & $\frac{1}{4}\left(J_{z}+\sqrt{B_{-}}\right)$\tabularnewline
\hline 
10  & $\frac{1}{4}\left(J_{z}-\sqrt{J^{2}+4J_{x}^{2}}\right)$  & $\frac{1}{4}\left(J_{z}-\sqrt{B_{-}}\right)$\tabularnewline
\hline 
11  & $\frac{1}{4}\left(J_{z}+\sqrt{J^{2}+16J_{x}^{2}}\right)$  & $2\sqrt{Q}\cos\left(\frac{\theta_{+}}{3}\right)-\frac{1}{12}\left(3J-J_{z}\right)$\tabularnewline
\hline 
12  & $\frac{1}{4}\left(J_{z}-\sqrt{J^{2}+16J_{x}^{2}}\right)$  & $2\sqrt{Q}\cos\left(\frac{\theta_{+}-2\pi}{3}\right)-\frac{1}{12}\left(3J-J_{z}\right)$\tabularnewline
\hline 
13  & $\frac{1}{4}\left(J_{z}-2J+\sqrt{A_{-}}\right)$  & $2\sqrt{Q}\cos\left(\frac{\theta_{+}-4\pi}{3}\right)-\frac{1}{12}\left(3J-J_{z}\right)$\tabularnewline
\hline 
14  & $\frac{1}{4}\left(J_{z}-2J-\sqrt{A_{-}}\right)$  & $2\sqrt{Q}\cos\left(\frac{\theta_{-}}{3}\right)+\frac{1}{12}\left(3J+J_{z}\right)$\tabularnewline
\hline 
15  & $\frac{1}{4}\left(J_{z}+2J+\sqrt{A_{+}}\right)$  & $2\sqrt{Q}\cos\left(\frac{\theta_{-}-2\pi}{3}\right)+\frac{1}{12}\left(3J+J_{z}\right)$\tabularnewline
\hline 
16  & $\frac{1}{4}\left(J_{z}+2J-\sqrt{A_{+}}\right)$  & $2\sqrt{Q}\cos\left(\frac{\theta_{-}-4\pi}{3}\right)+\frac{1}{12}\left(3J+J_{z}\right)$\tabularnewline
\hline 
\end{tabular}

\caption{\label{tab:Heisenberg-spin-eigenvalues,}Eigenvalues of spin-1/2 Heisenberg
decorated triangles, in which the coefficients are given in Eqs.(\ref{eq:A}-\ref{eq:theta}).
The eigenvalues $\{\varepsilon_{1}^{(3)},\varepsilon_{2}^{(3)},\varepsilon_{14}^{(3)},\varepsilon_{16}^{(3)},\varepsilon_{8}^{(1)}\}$
are the originator of the ground state energy of the spin-1/2 Ising-XXZ
model in the decorated honeycomb lattice.}
\end{table}

On the other hand, let us assume that the Hamiltonian of spin-1/2
Ising model on the effective triangular lattice can be expressed as
follows
\begin{equation}
\mathcal{H}_{\mathrm{eff}}=\tilde{\mathcal{H}}=-K_{0}-K\sum_{<i,j>}s_{i}s_{j},\label{eq:H-eff}
\end{equation}
in which $K_{0}$ corresponds to an effective ``constant'' energy,
and $K$ is the effective coupling parameter of Ising model on the
triangular lattice, while the summation is performed over the nearest-neighbor
interactions. So, the Boltzmann factor per each unit cell is given
by
\begin{equation}
\tilde{w}(\{s_{1},s_{2},s_{3}\})={\rm e}^{\beta K_{0}+\beta K(s_{1}s_{2}+s_{2}s_{3}+s_{3}s_{1})}.
\end{equation}

The spin-1/2 Ising model on the effective triangular lattice has only
two configurations, $\{+++\}$ and $\{++-\}$ (or its spin inversion),
corresponding to $\varsigma=3/2$ and $\varsigma=1/2$, respectively.
Therefore, both models should be equivalent, and the Boltzmann factor
for both lattices must satisfy the condition $\tilde{w}(\varsigma)=w(\varsigma)$.
After carrying out a direct star-triangle transformation\citep{Roj-sou11},
we get a couple of algebraic equations with two unknown parameters
$K_{0}$ and $K$ to be determined, so the algebraic system equations
read as follows
\begin{align}
\tilde{w}_{1}=\mathrm{e}^{\beta K_{0}}\exp\left(-\tfrac{\beta K}{4}\right)= & w_{1},\\
\tilde{w}_{3}=\mathrm{e}^{\beta K_{0}}\exp\left(\tfrac{3\beta K}{4}\right)= & w_{3}.
\end{align}

Subsequently, the unknown parameters of the effective triangular Ising
model could be expressed in terms of arbitrary parameters of the spin-1/2
Ising-XXZ model on a decorated honeycomb lattice {[}see Eq. \eqref{eq:Ham-1}{]}
\begin{align}
K= & \frac{1}{\beta}\ln\left(\tfrac{w_{3}}{w_{1}}\right),\label{eq:3-legs-K}\\
K_{0}= & \frac{1}{4\beta}\ln\left(w_{1}^{3}w_{3}\right),\label{eq:3-legs-K0}
\end{align}
where $w_{1}$ and $w_{3}$ are given by \eqref{eq:w(3/2)} and \eqref{eq:w(1/2)}
respectively. 

\subsection{Zero temperature Phase diagram}

\begin{figure}
\includegraphics[scale=0.51]{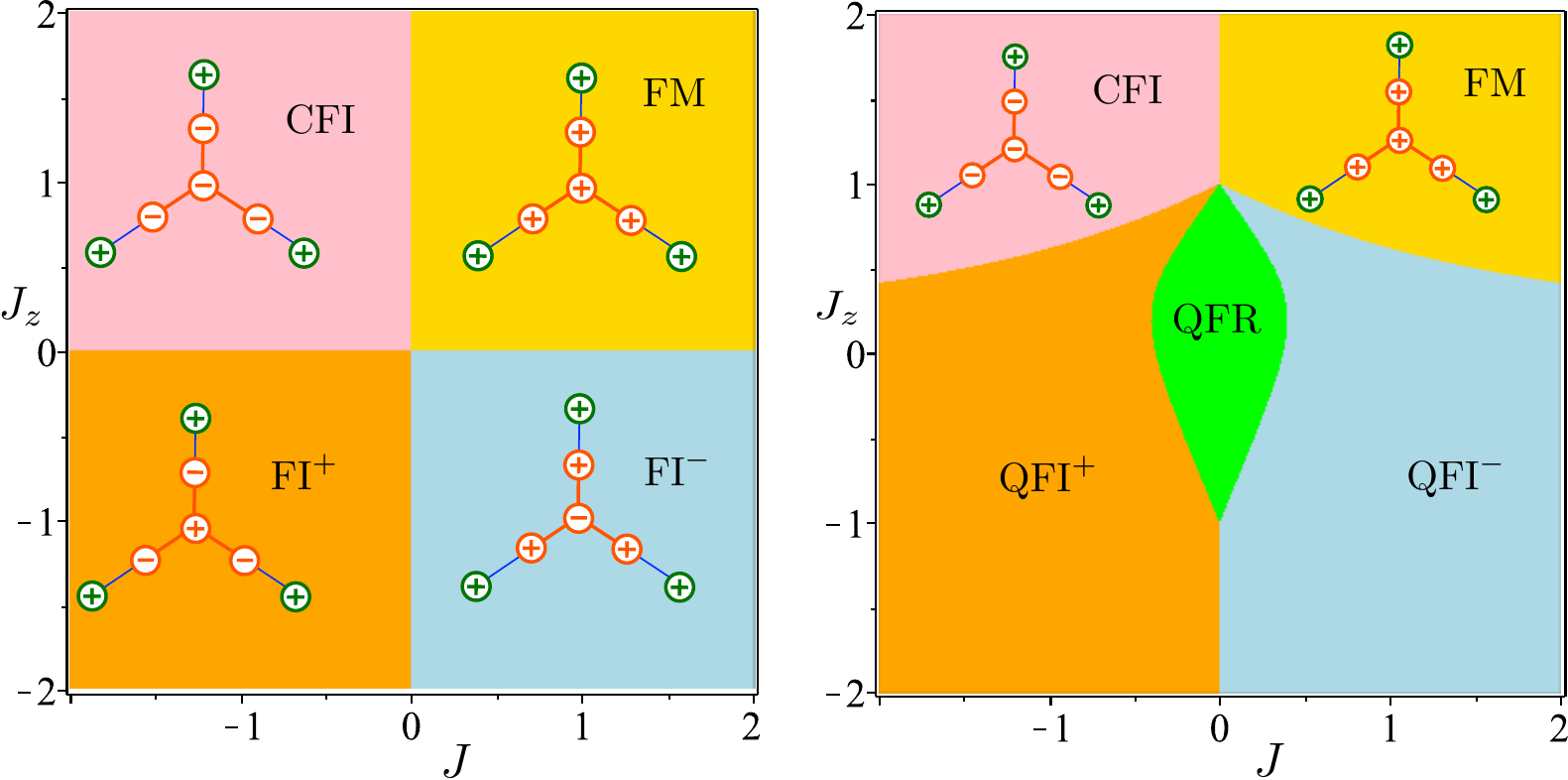}\caption{\label{fig: 0T phD}Zero temperature phase diagram in the plane $J-J_{z}$.
(a) For Ising model on a decorated honeycomb lattice ($J_{x}=0$).
(b) For Ising-XXZ model on a decorated honeycomb lattice, assuming
$J_{x}=1$.}
\end{figure}

Before exploring the finite temperature features, let us first study
the zero-temperature phase transition when $xy$-component exchange
interaction $J_{x}$ is taken into account. In Fig.\ref{fig: 0T phD}
is illustrated the zero temperature phase diagram in the plane $J-J_{z}$.
In panel (a), we report a typical phase diagram of the spin-1/2 Ising
model on a decorated honeycomb lattice ($J_{x}=0$) \cite{syozi}.
That is, we report one ferromagnetic ($\mathrm{FM}$) phase for $J>0$
and $J_{z}>0$, another classical ferrimagnetic ($\mathrm{CFI}$)
phase for $J<0$ and $J_{z}>0$, and two types of ferrimagnetic ($\mathrm{FI}^{\pm}$)
phases for $J_{z}<0$ with $J<0$ and $J>0$, respectively.

Already in Fig.\ref{fig: 0T phD}b is depicted the phase diagram for
an Ising-XXZ model on a decorated honeycomb lattice, described by
the Hamiltonian \eqref{eq:Ham-1}. The ground state energy comes from
the eigenvalues $\{\varepsilon_{1}^{(3)},\varepsilon_{2}^{(3)},\varepsilon_{14}^{(3)},\varepsilon_{16}^{(3)},\varepsilon_{8}^{(1)}\}$
of Table \ref{tab:Heisenberg-spin-eigenvalues,}. First, we express
the ground state energy per unit cell for ferromagnetic ($\mathrm{FM}$)
phase
\begin{equation}
E_{\mathrm{FM}}=-\frac{3}{4}\left(J_{z}+J\right),
\end{equation}
and the corresponding ground state per unit cell is
\begin{equation}
|\mathrm{FM}\rangle\mapsto\bigl|{}_{+}\overset{+}{\ocircle}_{+}\bigr\rangle\otimes\bigl|\overset{+}{_{+}+_{+}}\bigr\rangle.
\end{equation}
 In the direct product, the left ket corresponds to Ising spin configurations,
while the second ket corresponds to the Heisenberg spin configurations.
Indeed, the FM state is invariant under full spin inversion, so this
state is two-fold degenerate.

Let us define the magnetization for each configuration as follow:
$m^{I}=0.5$ for Ising spin, $m_{0}^{H}=0.5$ for the inner Heisenberg
spin, and $m_{1}^{H}=(0.5+0.5+0.5)/3=0.5$ for the Heisenberg spin
that connects to Ising spin. Likewise, the total magnetization per
unit cell may be obtained as follows
\begin{equation}
m_{t}=\frac{m^{I}+m_{0}^{H}+3m_{1}^{H}}{5}.
\end{equation}
Thus, for the ferromagnetic case, the total magnetization obviously
becomes $m_{t}=0.5$.

Second, the phase diagram in Fig.\ref{fig: 0T phD}b also shows a
classical ferrimagnetic ($\mathrm{CFI}$) phase per unit cell for
$J>0$ and $J_{z}<0$, providing the result in
\begin{equation}
E_{\mathrm{CFI}}=-\frac{3}{4}\left(J_{z}-J\right).
\end{equation}
Here, we call this a ``classical'' phase because the ground state
is independent of the parameter $J_{x}$. Thus the ground state per
unit cell of the aforementioned energy becomes
\begin{equation}
|\mathrm{CFI}\rangle\mapsto\bigl|{}_{-}\overset{-}{\ocircle}_{-}\bigr\rangle\otimes\bigl|\overset{+}{_{+}+_{+}}\bigr\rangle.
\end{equation}
The left ket corresponds to the Ising spin setups in the direct product,
whereas the second ket corresponds to the Heisenberg spin setups.
Obviously, the model is invariant under global spin inversion, so
the CFI state is two-fold degenerate.

Note that the magnetizations are $m^{I}=-0.5$, $m_{0}^{H}=0.5$,
$m_{1}^{H}=0.5$, and $m_{t}=0.3$.

Furthermore, the third phase corresponds to a quantum ferrimagnetic
($\mathrm{QFI}^{-}$) phase whose ground state energy per unit cell
reads
\begin{equation}
E_{\mathrm{QFI^{-}}}=\frac{1}{4}\left(J_{z}-2J-\sqrt{\left(J-2J_{z}\right)^{2}+12J_{x}^{2}}\right).
\end{equation}
And corresponding ground states per unit cell is given by
\begin{equation}
|\mathrm{QFI^{-}}\rangle\mapsto\tfrac{\bigl|{}_{+}\overset{+}{\ocircle}_{+}\bigr\rangle\otimes\left[J_{x}\bigl|\bigl(\overset{+}{_{+}+_{-}}\bigr)\bigr\rangle+\varepsilon_{14}^{(3)}\bigl|\overset{+}{_{+}-_{+}}\bigr\rangle\right]}{\sqrt{3J_{x}^{2}+(\varepsilon_{14}^{(3)})^{2}}},
\end{equation}
in direct product, the left ket corresponds to the Ising spins, whereas
right ket denotes the Heisenberg spins, with $\bigl|\bigl(\overset{+}{_{+}s_{-}}\bigr)\bigr\rangle\equiv\bigl\{\bigl|\overset{+}{_{+}s_{-}}\bigr\rangle+\bigl|\overset{-}{_{+}s_{+}}\bigr\rangle+\bigl|\overset{+}{_{-}s_{+}}\bigr\rangle\bigr\}$.
Here, we call this ``quantum'' phase because the ground state and
the corresponding phase depend on $J_{x}$. Since the system is invariant
under global spin inversion, the $\mathrm{QFI}^{-}$ state is also
two-fold degenerate. 

In this case the magnetization can be recovered using the Eq.\eqref{eq:M_h-t},
and taking the limit of $T=0$, which results in
\begin{alignat}{1}
m^{I}= & \frac{1}{2},\\
m_{0}^{H}= & -\frac{J-2J_{z}}{2\sqrt{\left(J-2J_{z}\right)^{2}+12J_{x}^{2}}},\\
m_{1}^{H}= & \frac{J-2J_{z}}{6\sqrt{\left(J-2J_{z}\right)^{2}+12J_{x}^{2}}}+\frac{1}{3},\\
m_{t}= & \frac{3}{10}.
\end{alignat}

However, the magnetization of all Heisenberg spins is constant ($m_{0}^{H}+3m_{1}^{H}=1$),
independent of Hamiltonian parameters. 

The other quantum ferrimagnetic ($\mathrm{QFI}^{+}$) phase arises
for $J<0$ and $J_{z}<0$, whose ground state energy per unit cell
results in
\begin{equation}
E_{\mathrm{QFI^{+}}}=\frac{1}{4}\left(J_{z}+2J-\sqrt{\left(J+2J_{z}\right)^{2}+12J_{x}^{2}}\right).
\end{equation}
Whereas the ground state $\mathrm{QFI}^{+}$ per unit cell becomes

\begin{equation}
|\mathrm{QFI^{+}}\rangle\mapsto\tfrac{\bigl|{}_{-}\overset{-}{\ocircle}_{-}\bigr\rangle\otimes\left[J_{x}\bigl|\bigl(\overset{+}{_{+}+_{-}}\bigr)\bigr\rangle+\varepsilon_{16}^{(3)}\bigl|\overset{+}{_{+}-_{+}}\bigr\rangle\right]}{\sqrt{3J_{x}^{2}+(\varepsilon_{16}^{(3)})^{2}}},
\end{equation}
in the direct product, the left ket denotes Ising spins, and the right
spins correspond to the Heisenberg spins. Similar to previous states,
the $\mathrm{QFI}^{+}$is invariant upon reversing all spins, so the
$\mathrm{QFI^{+}}$ state is again two-fold degenerate. 

Again using Eq.\eqref{eq:M_h-t} and taking the limit $T=0$, one
can obtain straightforwardly the following magnetizations
\begin{alignat}{1}
m^{I}= & -\frac{1}{2},\\
m_{0}^{H}= & \frac{J+2J_{z}}{2\sqrt{\left(J+2J_{z}\right)^{2}+12J_{x}^{2}}},\\
m_{1}^{H}= & -\frac{J+2J_{z}}{6\sqrt{\left(J+2J_{z}\right)^{2}+12J_{x}^{2}}}+\frac{1}{3},\\
m_{t}= & \frac{1}{10}.
\end{alignat}
Analogously to the $\mathrm{QFI^{-}}$ phase, the magnetization of
all Heisenberg spins becomes $m_{0}^{H}+3m_{1}^{H}=1$, independent
of the Hamiltonian parameters.

The last phase arises due to including the $xy$-component exchange
interaction $J_{x}$. This state is a quantum spin frustrated ($\mathrm{QFR}$)
phase, whose ground state energy per unit cell becomes from Table
\ref{tab:Heisenberg-spin-eigenvalues,} by
\begin{equation}
E_{\mathrm{QFR}}=\frac{1}{4}\left(J_{z}-\sqrt{B_{+}}\right),\label{eq:QFR}
\end{equation}
with $B_{+}$ being defined in Eq.\eqref{eq:B}. After a bit of cumbersome
algebraic manipulation, the corresponding eigenstate per unit cell
results in

\begin{equation}
|\mathrm{QFR\rangle}=\bigl|{}_{+}\overset{?}{\ocircle}_{-}\bigr\rangle\otimes|hAF_{?}\rangle
\end{equation}
where $?=\pm$, and 
\begin{alignat}{1}
|hAF_{+}\rangle\mapsto & \tfrac{1}{\mathcal{N}_{+}}\left[a_{2}\bigl|\overset{-}{_{-}+_{+}}\bigr\rangle+a_{1}\bigl|\overset{-}{_{+}+_{-}}\bigr\rangle+a_{1}\bigl|\overset{+}{_{-}+_{-}}\bigr\rangle\right.\nonumber \\
 & +\left.b_{1}\bigl|\overset{-}{_{+}-_{+}}\bigr\rangle+b_{1}\bigl|\overset{+}{_{-}-_{+}}\bigr\rangle+b_{2}\bigl|\overset{+}{_{+}-_{-}}\bigr\rangle\right],\\
|hAF_{-}\rangle\mapsto & \tfrac{1}{\mathcal{N}_{-}}\left[\bar{a}_{1}\bigl|\overset{-}{_{-}+_{+}}\bigr\rangle+\bar{a}_{2}\bigl|\overset{-}{_{+}+_{-}}\bigr\rangle+\bar{a}_{1}\bigl|\overset{+}{_{-}+_{-}}\bigr\rangle\right.\nonumber \\
 & +\left.\bar{b}_{1}\bigl|\overset{-}{_{+}-_{+}}\bigr\rangle+\bar{b}_{2}\bigl|\overset{+}{_{-}-_{+}}\bigr\rangle+\bar{b}_{1}\bigl|\overset{+}{_{+}-_{-}}\bigr\rangle\right],
\end{alignat}
with $\mathcal{N}_{+}=\sqrt{2a_{1}^{2}+a_{2}^{2}+2b_{1}^{2}+b_{2}^{2}}$,
$\mathcal{N}_{-}=\sqrt{2\bar{a}_{1}^{2}+\bar{a}_{2}^{2}+2\bar{b}_{1}^{2}+\bar{b}_{2}^{2}}$
being the normalization coefficient. Whereas the Heisenberg spins
state $|hAF_{+}\rangle$ coefficients are given by 
\begin{alignat}{1}
a_{1}= & J_{x}\left(\sqrt{B_{+}}-3J\right)\left(10J^{2}+J_{x}^{2}-C^{2}\right)+\frac{2J}{J_{x}}b_{1},\nonumber \\
a_{2}= & J_{x}\left(\sqrt{B_{+}}-3J\right)\left(C^{2}+2J^{2}-5J_{x}^{2}\right),\nonumber \\
b_{2}= & 4\left(3J^{2}-J_{x}^{2}\right)\left(C^{2}+2J^{2}-J_{x}^{2}\right)+\frac{2J}{J_{x}}a_{2},\nonumber \\
b_{1}= & 8J_{x}^{2}\left(3J^{2}-J_{x}^{2}\right),\label{eq:gen-coeff}
\end{alignat}
where $C^{2}=\sqrt{(2J^{2}+3J_{x}^{2})^{2}+16J^{2}J_{x}^{2}}$, and
for $|hAF_{-}\rangle$ becomes

\begin{alignat}{1}
\bar{a}_{1}= & J_{x}\left(\sqrt{B_{+}}+3J\right)\left(10J^{2}+J_{x}^{2}-C^{2}\right)-\frac{2J}{J_{x}}\bar{b}_{1},\nonumber \\
\bar{a}_{2}= & J_{x}\left(\sqrt{B_{+}}+3J\right)\left(C^{2}+2J^{2}-5J_{x}^{2}\right),\nonumber \\
\bar{b}_{2}= & 4\left(3J^{2}-J_{x}^{2}\right)\left(C^{2}+2J^{2}-J_{x}^{2}\right)-\frac{2J}{J_{x}}\bar{a}_{2},\nonumber \\
\bar{b}_{1}= & 8J_{x}^{2}\left(3J^{2}-J_{x}^{2}\right).\label{eq:gen-coeff-1}
\end{alignat}
 It should be noted that if $J\mapsto-J$, the following property
must held $a_{1}\leftrightarrows b_{1}$ and $a_{2}\leftrightarrows b_{2}$,
and equivalently for $\bar{a}_{1}\leftrightarrows\bar{b}_{1}$ and
$\bar{a}_{2}\leftrightarrows\bar{b}_{2}$.

Note that the magnetization $m^{I}=0$, $m_{0}^{H}=0$, and $m_{1}^{H}=0$,
then $m_{t}=0$. This result could be confirmed from Eqs.\eqref{eq:s0}
and \eqref{eq:M_h-t}. 

\begin{figure}
\includegraphics[scale=0.7]{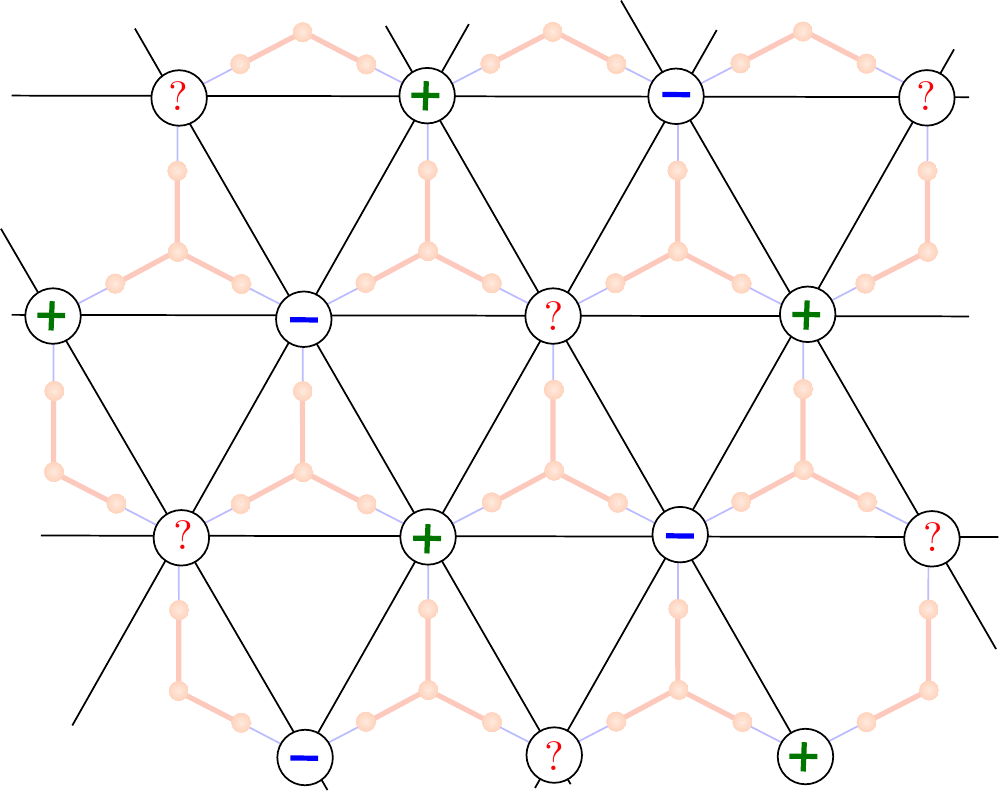}\caption{\label{fig:AFM-fru}A schematic arrangement of spin frustration in
the effective triangular lattice.}
\end{figure}

In Fig.\ref{fig:AFM-fru} is illustrated a frustrated Ising spin arrangement
for the spin-1/2 Ising model on the triangular lattice. In this arrangement,
two vertices of the triangle have two opposite spins, and the third
spin must take at random upward or downward. We can also check that
six spins surround each random spin by alternating fixed spins.

Indeed, the probability amplitude of the Heisenberg spins state is
somewhat cumbersome to be inspected quickly, so let us analyze some
limiting cases: 

(i) First, we consider a trivial case when $J=0$, so the QFR state
reduces to 
\begin{alignat}{1}
|\mathrm{QFR}\rangle\mapsto & \tfrac{1}{\sqrt{6}}\bigl|{}_{+}\overset{?}{\ocircle}_{-}\bigr\rangle\otimes\left[\bigl|\overset{-}{_{-}+_{+}}\bigr\rangle+\bigl|\overset{-}{_{+}+_{-}}\bigr\rangle+\bigl|\overset{+}{_{-}+_{-}}\bigr\rangle\right.\nonumber \\
 & +\left.\bigl|\overset{-}{_{+}-_{+}}\bigr\rangle+\bigl|\overset{+}{_{-}-_{+}}\bigr\rangle+\bigl|\overset{+}{_{+}-_{-}}\bigr\rangle\right],
\end{alignat}
 where each spin configuration are equally probably, since there is
no exchange interaction with Ising spins. 

(ii) The second case we consider is when $J>0$ and $J_{x}\rightarrow0$,
then for this purpose, we take the Taylor expansion in \eqref{eq:gen-coeff}
around $J_{x}=0$ up to order $\mathcal{O}(J_{x}^{4})$. Therefore
the coefficients simply become

\begin{alignat}{3}
a_{1}= & \tfrac{J_{x}}{J}-\tfrac{5}{2}\left(\tfrac{J_{x}}{J}\right)^{3}, & \quad & \quad & \bar{a}_{1}= & -\tfrac{J_{x}}{J}-\tfrac{5}{2}\left(\tfrac{J_{x}}{J}\right)^{3},\nonumber \\
a_{2}= & \tfrac{1}{3}\left(\tfrac{J_{x}}{J}\right)^{3}, &  &  & \bar{a}_{2}= & -\tfrac{1}{3}\left(\tfrac{J_{x}}{J}\right)^{3},\nonumber \\
b_{2}= & 1-\left(\tfrac{J_{x}}{J}\right)^{2}, &  &  & \bar{b}_{2}= & 1-\left(\tfrac{J_{x}}{J}\right)^{2},\nonumber \\
b_{1}= & \left(\tfrac{J_{x}}{J}\right)^{2}, &  &  & \bar{b}_{1}= & \left(\tfrac{J_{x}}{J}\right)^{2}.\label{eq:part-coeff}
\end{alignat}
It is obvious that the coefficients hold the aforementioned property
when $J\mapsto-J$.

Note that for exactly $J_{x}=0$ (Ising model on a decorated honeycomb
lattice\cite{syozi}), the only coefficient that survives is $b_{2}$,
thus the frustrated state reduces to $|\mathrm{QFR}\rangle\mapsto\bigl|{}_{+}\overset{+}{\ocircle}_{-}\bigr\rangle\otimes\bigl|\overset{+}{_{-}-_{+}}\bigr\rangle$
or $\bigl|{}_{+}\overset{-}{\ocircle}_{-}\bigr\rangle\otimes\bigl|\overset{-}{_{+}-_{+}}\bigr\rangle$,
but the eigenvalue \eqref{eq:QFR} is no longer the ground state energy,
unless in a trivial case when $J_{z}=0$. So that the quantum spin
frustrated state vanishes, thereby confirming the absence of quantum
spin frustration of the spin-1/2 Ising model on a decorated honeycomb
lattice\cite{syozi}.

Another thing we to point out is the absence of symmetry under local
Heisenberg spin reversal. Since $a_{1}\ne b_{1}$ and $a_{2}\ne b_{2}$,
this is already manifest in the coefficients provided in \eqref{eq:gen-coeff},
which is even more evident in the amplitude probability of Eqs. \eqref{eq:part-coeff}. 

It is worth remarking that the $\mathrm{QFR}$ phase exclusively arises
due to the $J_{x}$ contribution. This means that in the effective
triangular lattice, the $J_{x}$ is responsible for generating the
non-fitting antiferromagnetic arrangement on triangular lattice of
the Ising spins\cite{wannier}. Once $1/3$ of Ising spins are set
to point down, other $1/3$ Ising spins are set to point up, while
the remaining $1/3$ must point up or down randomly as depicted in
fig.\ref{fig:AFM-fru}. This configuration gains a weight of $2^{1/3}$
per spin, and in addition, there is still a considerable amount of
contingent freedom to be taken into account. Because random spins
can vary independently; thus three random spins forming a triangle
could often occur with the same signs (for detail, see Ref \cite{wannier}). 

\begin{figure}
\includegraphics[scale=0.6]{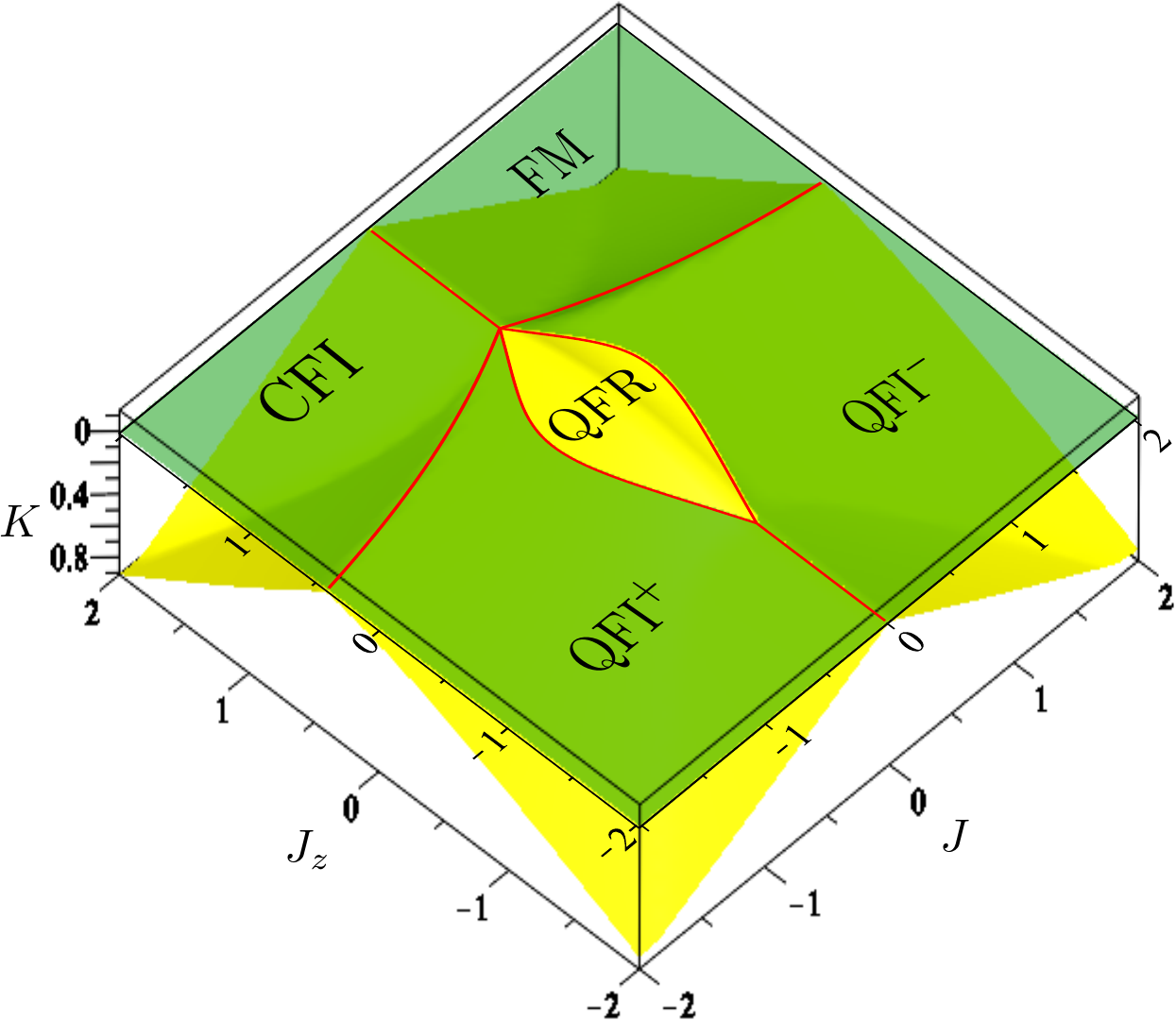}

\caption{\label{fig:Keff}The effective interaction $K$ given by \eqref{eq:3-legs-K}
by a yellow surface as a function of $J$ and $J_{z}$, assuming fixed
$J_{x}=1$. Transparent green plane corresponds to $K=0$, the negative
$K$ is illustrated on the top of surface for convenience.}
\end{figure}

It is also important to note that the phase boundary curves described
in Figs.\ref{fig: 0T phD}b and \ref{fig:Keff} are reported by 
\begin{alignat}{1}
\mathrm{QFI}^{+}-\mathrm{QFR},\;\rightarrow\; & J_{z}=\tfrac{-J\pm\sqrt{(2J+\sqrt{B_{+}})^{2}-12J_{x}^{2}}}{2},\\
\mathrm{QFI}^{-}-\mathrm{QFR},\;\rightarrow\; & J_{z}=\tfrac{J\pm\sqrt{(2J-\sqrt{B_{+}})^{2}-12J_{x}^{2}}}{2},\\
\mathrm{QFI}^{-}-\mathrm{FM},\;\rightarrow\; & J_{z}=\tfrac{\sqrt{J^{2}+4J_{x}^{2}}-J}{4},\\
\mathrm{QFI}^{+}-\mathrm{FM},\;\rightarrow\; & J_{z}=\tfrac{\sqrt{J^{2}+4J_{x}^{2}}+J}{4}.
\end{alignat}

In fig.\ref{fig:Keff} the effective parameter $K$ is illustrated
as a yellow surface in the plane $J-J_{z}$ for a fixed $J_{x}=1$
and at zero temperature. Just for convenience, in order to show a
slight negative $K$ on the top of the surface, the parameter $K$
increases while going down, so confirming the existence of a frustrated
region. At the same time, the transparent green flat region corresponds
to $K=0$. 

We leave the analysis of residual entropy of the frustrated region,
to next section, because to obtain this result, we need to use the
free energy at finite temperature.

\section{Thermodynamics}

Now we are in a position to discuss the thermodynamics of the present
model. So the first thing to do is to get the free energy per unit
cell of the model, which could be written as
\begin{equation}
f_{\mathrm{dh}}=-K_{0}+\tilde{f}_{\mathrm{t}},
\end{equation}
where $f_{\mathrm{t}}$ denotes the free energy of the Ising model
in the effective triangular lattice \citep{syozi,kano53,azaria-prl87},
and $K_{0}$ can be understood as an effective \textquotedbl constant\textquotedbl{}
energy of the effective triangular lattice, since it is independent
of spins on a triangular lattice but depends of the temperature.

Consequently, the free energy of effective triangular lattice per
unit cell \citep{syozi,kano53,azaria-prl87} could be expressed using
a single integral \cite{fan-wu-70,stephenson}, as follows
\begin{equation}
f_{\mathrm{t}}=\frac{T}{4}\ln(r)-\frac{T}{4\pi}\int_{-\pi}^{\pi}\ln\left[\mathcal{A}(\phi)+\sqrt{\mathcal{Q}(\phi)}\right]\mathrm{d}\phi,
\end{equation}
where $\mathcal{A}(\phi)$ and $\mathcal{Q}(\phi)$ are defined by
\begin{alignat}{1}
\mathcal{A}(\phi)= & \tfrac{1}{2}r^{2}+\tfrac{3}{2}+(1-r)\cos(\phi),\label{eq:A-p}\\
\mathcal{Q}(\phi)= & \mathcal{A}(\phi)^{2}-2(r-1)^{2}\left[1+\cos(\phi)\right],\label{eq:Q-p}
\end{alignat}
with $r=\frac{w_{3}}{w_{1}}.$ Thus, the free energy per unit cell
of decorated triangular lattice becomes
\begin{equation}
f_{\mathrm{dh}}=-T\ln\left(w_{1}\right)-\frac{T}{4\pi}\int_{-\pi}^{\pi}\ln\left[\mathcal{A}(\phi)+\sqrt{\mathcal{Q}(\phi)}\right]\mathrm{d}\phi.\label{eq:fek}
\end{equation}
Alternatively, let us define the following quantity
\begin{equation}
u=\frac{2(1-r)}{r^{2}+3},\quad\text{where}\quad-\tfrac{1}{3}<u\leqslant\tfrac{2}{3}.
\end{equation}
Using this relation, the free energy can be re-expressed as follows
\begin{alignat}{1}
f_{\mathrm{dh}}=- & \frac{T}{2}\ln\left(\tfrac{w_{3}^{2}+3w_{1}^{2}}{2}\right)-\frac{T}{4}\left\{ \ln\left[\tfrac{2-u^{2}}{4}+\tfrac{\sqrt{1-u^{2}}}{2}\right]\right\} \nonumber \\
 & -\frac{T}{4\pi}\int_{-\pi}^{\pi}\ln\left[1+\sqrt{1-\kappa(\phi)^{2}}\right]{\rm d}\phi,
\end{alignat}
where
\begin{equation}
\kappa(\phi)=\frac{u\sqrt{2\left(1+\cos(\phi)\right)}}{\left(1+u\cos(\phi)\right)}.
\end{equation}

Before further investigation concerning thermodynamic properties,
two interesting cases should be noted.

First, we analyze the residual entropy in the $\mathrm{QFR}$ region,
which occurs when $w_{3}<w_{1}$ and assuming $T\rightarrow0$, this
implies that $r=\frac{w_{3}}{w_{1}}\rightarrow0$, then the Eqs.\eqref{eq:A-p}
and \eqref{eq:Q-p} reduce to
\begin{alignat}{1}
\mathcal{A}(\phi)= & \cos(\phi)+\tfrac{3}{2},\label{eq:Ap}\\
\mathcal{Q}(\phi)= & \left[\cos(\phi)+\tfrac{3}{2}\right]^{2}-2\left[1+\cos(\phi)\right],\label{eq:Qp}
\end{alignat}
both functions are independent of $T$ and $r$, then after some algebraic
manipulation, the residual entropy results in
\begin{alignat}{1}
\mathcal{S}= & \tfrac{1}{3}\ln(2)+\frac{1}{4\pi}\int_{-\frac{2\pi}{3}}^{\frac{2\pi}{3}}\ln\left(\cos(\phi)+1\right){\rm d}\phi,\nonumber \\
= & \frac{2}{\pi}\int_{0}^{\frac{\pi}{3}}\ln\left(2\cos(\phi)\right){\rm d}\phi,\nonumber \\
= & 0.32306594721945.
\end{alignat}

As mentioned above, the residual entropy arises because of the contribution
of the quantum exchange interaction $J_{x}$, turning the effective
$K$ of triangular lattice becoming antiferromagnetic ($K<0$) since
$r\rightarrow0$, which implies that the model exhibits a frustrated
state. This result is precisely the same residual entropy found by
Wannier\cite{wannier} for the non-fitting antiferromagnetic lattice. 

Second, we study the residual entropy when $\frac{w_{3}}{w_{1}}=r=1$
and $T\rightarrow0$. In this case, we have $A(\phi)=2$ and $\mathcal{Q}(\phi)=2^{2}$,
so the free energy merely reduces to
\begin{equation}
f_{\mathrm{dh}}=-T\ln\left(w_{1}\right)-\frac{T}{4\pi}\int_{-\pi}^{\pi}\ln\left(2^{2}\right)\mathrm{d}\phi,\label{eq:fek-1}
\end{equation}
implying that residual entropy becomes $\mathcal{S}=\ln(2)$. This
residual entropy emerges in the straight line given by $J=0$ and
in the phase boundary between $\mathrm{QFR}-\mathrm{QFI^{\pm}}$,
illustrated in Fig.\ref{fig:Keff} by red curves.

It is easy to convince that for $\mathrm{QFI}^{\pm}$, CFI and FM
states we have $w_{3}>w_{1}$ and assuming $T\rightarrow0$, which
implies that $r=\frac{w_{3}}{w_{1}}\rightarrow\infty$, thus as expected
the corresponding residual entropy becomes null.

\begin{figure}
\includegraphics[scale=0.5]{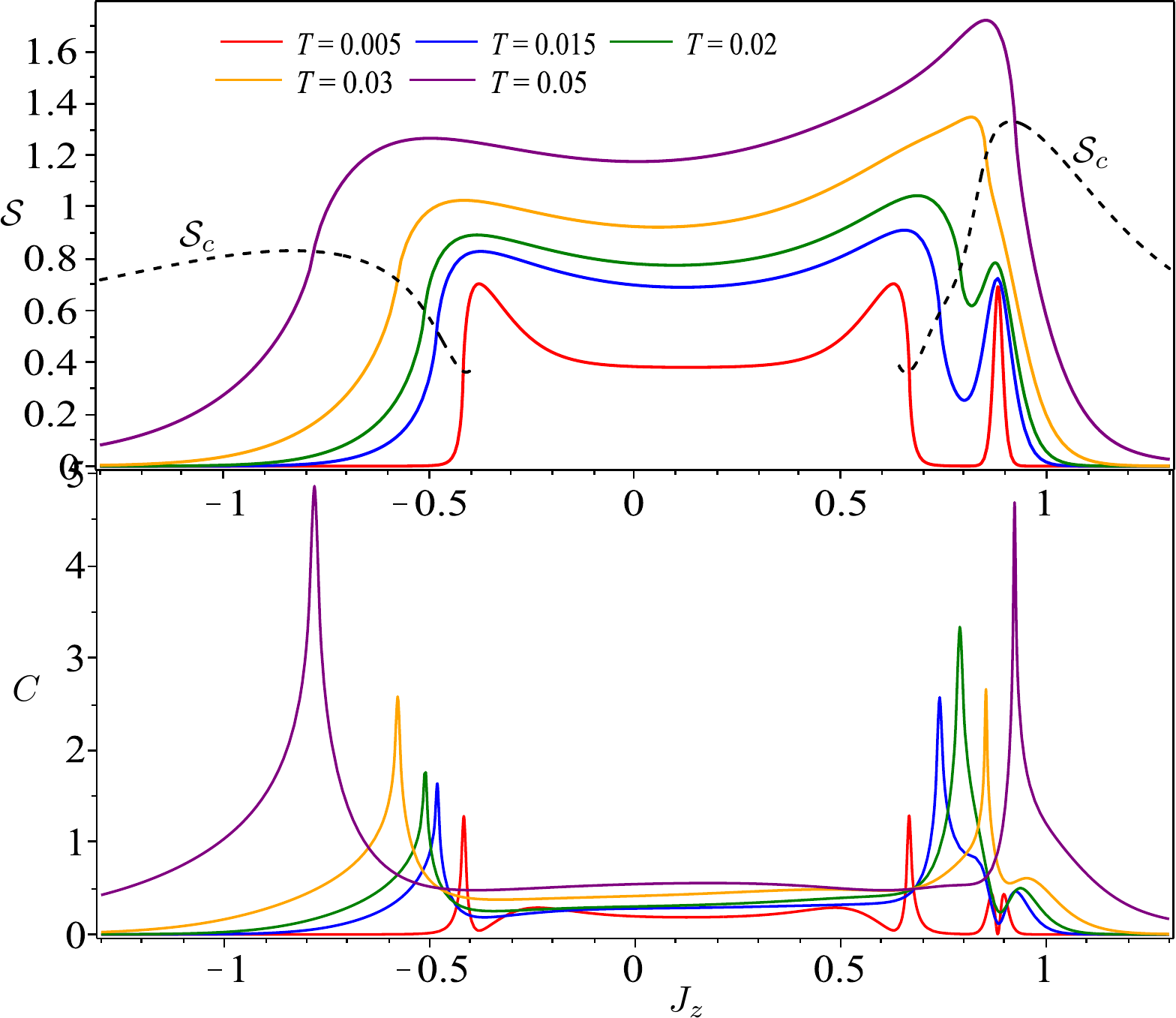}\caption{\label{fig:Entr-Csp-Jz}(Top) Entropy $\mathcal{S}$ as a function
of $J_{z}$, assuming $J_{x}=1$, $J=0.025$ and for fixed temperatures
$T=\{0.005,0.015,0.02,0.03,0.05\}$, dashed line corresponds to the
entropy at critical point. (Bottom) Specific heat as a function of
$J_{z}$ for the same set of parameters of the top panel.}
\end{figure}

Now, once we have free energy, we return to explore the thermodynamic
properties, it is possible to study several physical quantities of
the Ising-XXZ model on a decorated honeycomb lattice. Thus, in Fig.\ref{fig:Entr-Csp-Jz}(top)
is depicted the entropy as a function of $J_{z}$, assuming fixed
$J_{x}=1$ and $J=1$ for a number of fixed temperatures given inside
the panel. Particularly, let us start paying attention to the temperature
$T=0.005$; essentially, we observe three peaks at $J_{z}=-0.38$,
$J_{z}=0.63$ and $J_{z}=0.88$ due to the influence of phase transitions
at zero temperature in $\mathrm{QFI^{-}-QFR}$, $\mathrm{QFR}-\mathrm{QFI^{-}}$,
and $\mathrm{QFI^{-}-FM}$, respectively. Those peaks broaden as soon
as the temperature increases. In contrast, the dashed line corresponds
to the entropy $\mathcal{S}_{c}$ at the critical point as a function
of $J_{z}$. In Fig.\ref{fig:Entr-Csp-Jz}(bottom), we illustrate
the specific heat as a function of $J_{z}$, for the same set of entropy
parameters. Once again, let us focus at the low-temperature region
say $T=0.005$, the sharp \textquotedblleft peaks\textquotedblright{}
actually describes the logarithmic divergence due to the second order
phase transition. Meanwhile, the small double peaks that appear around
the phases $\mathrm{QFI^{-}-QFR}$ ($J_{z}=-0.38$), $\mathrm{QFR}-\mathrm{QFI^{-}}$
($J_{z}=0.63$), and $\mathrm{QFI^{-}-FM}$ ($J_{z}=0.88$), correspond
to the phase transition effect at zero temperature. Of course, these
anomalous peaks vanish as the temperature increases, surviving only
the logarithmic divergence at the critical temperature $T_{c}$.

\begin{figure}
\includegraphics[scale=0.5]{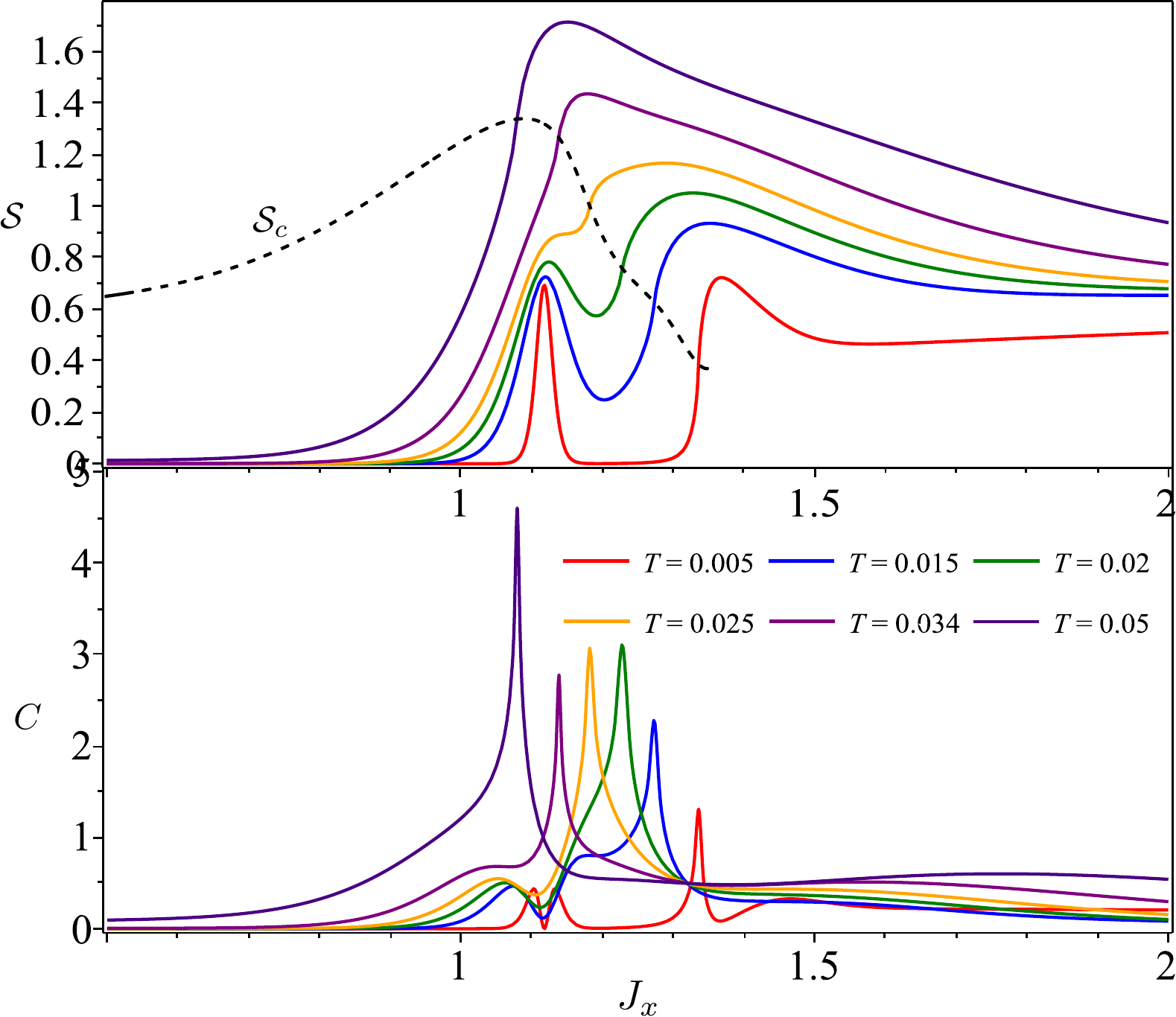}\caption{\label{fig:Entr-Csp-Jx}(Top) Entropy $\mathcal{S}$ as a function
of $J_{x}$, assuming $J_{z}=1$, $J=0.025$ and for fixed temperatures
$T=\{0.005,0.015,0.02,0.025,0.034,0.05\}$. (Bottom) Specific heat
as a function of $J_{x}$, for the same set of parameters of entropy.}
\end{figure}

In Fig.\ref{fig:Entr-Csp-Jx}(top) is reported the entropy as a function
of $J_{x}$, for fixed $J_{z}=1$, $J=0.025$ and assuming several
values of temperature $T=\{0.005,0.015,0.02,0.025,0.034,0.05\}$.
For temperature $T=0.005$, again, we observe two peaks, and the sharper
peak arises as a consequence of zero temperature phase transition
between $\mathrm{QFI^{-}-FM}$ ($J_{x}=1.118$), while the broader
peak indicates the influence of zero temperature phase transition
between $\mathrm{QFI^{-}-QFR}$ ($J_{x}=1.369$). When the temperature
increases, the influence of the zero-temperature phase transition,
shows that the peaks vanish. Meanwhile, the dashed line describes
entropy $\mathcal{S}_{c}$ at critical temperature as a function of
$J_{x}$. In Fig.\ref{fig:Entr-Csp-Jx}(bottom) is shown the specific
heat as a function of $J_{x}$ for the same set of parameters considered
for entropy. Although once more, for $T=0.005$ we observe a double
peak around the zero-temperature phase transition $\mathrm{QFI^{-}-FM}$.
This double peak disappears in the same way as in the previous case
when the temperature increases. We also observe at the critical temperature
a logarithmic divergence for each curve, corresponding to the second-order
phase transition.

\subsection{Critical temperature}

In the following, we will discuss one of the most significant properties
of 2D lattice models, the critical behavior at finite temperature.
It is well established that the critical temperature of the spin-1/2
Ising model on the triangular lattice is given by $\frac{K}{T_{c}}=\ln\left(3\right)$
\citep{syozi,barry}. Consequently, we can write the critical condition
as follows
\begin{equation}
\frac{w_{3}^{c}}{w_{1}^{c}}=r_{c}=3,\label{eq:crit-T}
\end{equation}
where $w_{1}^{c}$ and $w_{3}^{c}$ refer to the Boltzmann factors
at a critical temperature $T_{c}$, given by Eqs. \eqref{eq:w(3/2)}
and \eqref{eq:w(1/2)} at $T=T_{c}$, respectively. An explicit expression
involves a huge algebraic expression, which is irrelevant to write
explicitly here.

\begin{figure}
\includegraphics[scale=0.5]{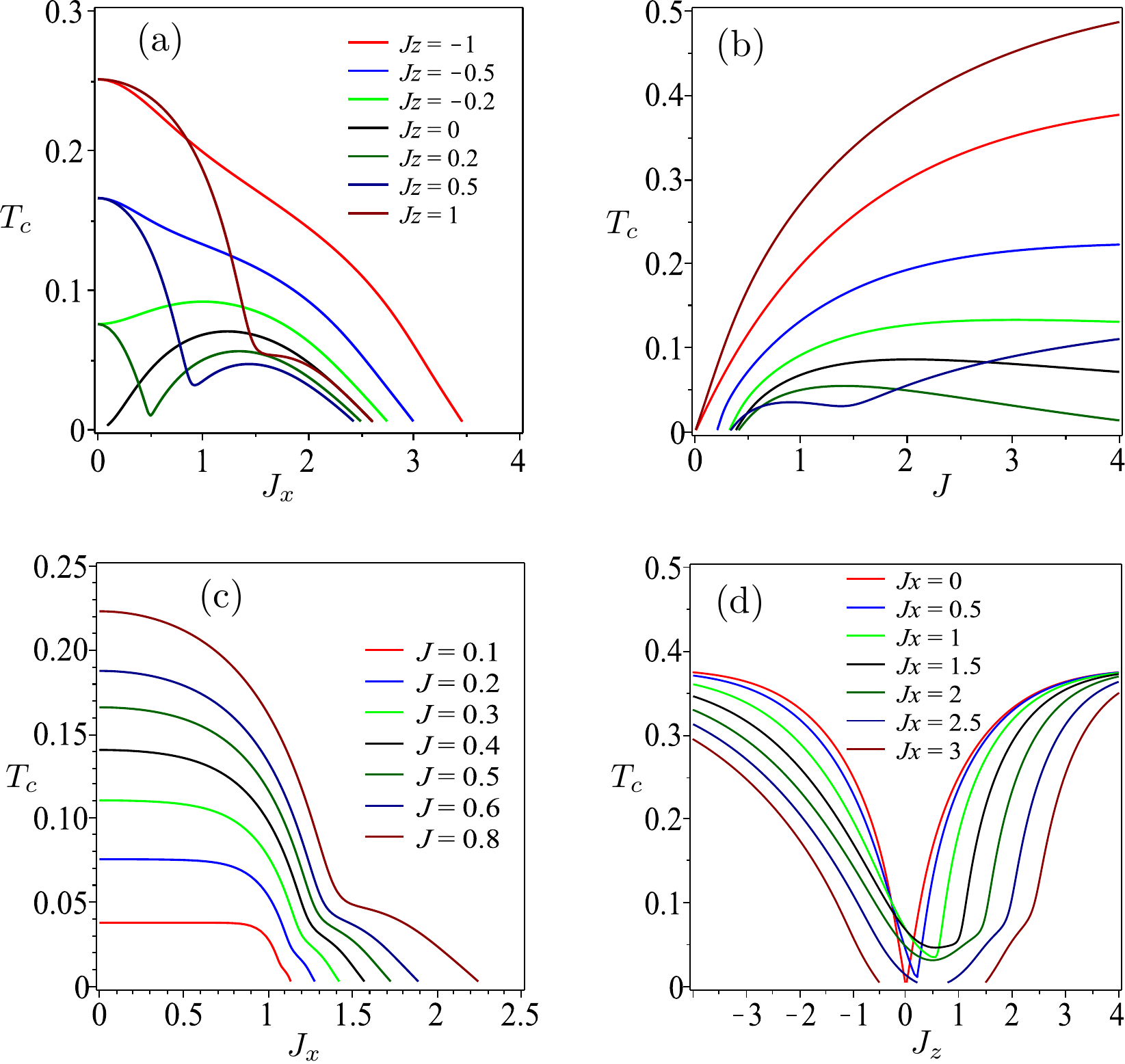}\caption{\label{fig:crit-temp}(a) Critical temperature $T_{c}$ as a function
of $J_{x}$, for fixed $J=1$ and certain values of $J_{z}$. (b)
Critical temperature $T_{c}$ as a function of $J$, for a range of
values in $J_{z}$ and fixed $J_{x}=1$. (c) Critical temperature
$T_{c}$ as a function of $J_{x}$, for several values of $J$ assuming
fixed $J_{z}=1$. (d) Critical temperature $T_{c}$ as a function
of $J_{z}$, for certain values of $J_{x}$ and fixed $J=1$. }
\end{figure}

In Fig.\ref{fig:crit-temp}a is illustrated the critical temperature
$T_{c}$ as a function of $J_{x}$ for several values of $J_{z}$
and fixing $J=1$. For $J_{z}>0$ we observe an interesting behavior
where appears a minimum, which corresponds to the interface between
$\mathrm{QFR}$ and $\mathrm{QFI^{\pm}}$, although for $J_{z}=0$
there is a special curve with two critical points at zero temperature
(for detail see Fig.\ref{fig: 0T phD}). For $|J_{z}|\gtrsim0.2$
only occurs one critical point at zero temperature. Furthermore, Fig.\ref{fig:crit-temp}b
depicts the critical temperature as a function of $J$, assuming $J_{x}=1$
and the same set of parameters for $J_{z}$ considered in panel (a).
Similarly, in panel (c), the critical temperature is depicted as a
function of $J_{z}$, keeping fixed $J_{z}=1$, and for a range of
values of $J$ as displayed inside the panel. Finally, panel (d) illustrates
the critical temperature as a function of $J_{z}$ assuming fixed
$J=1$ and for a set of values of $J_{x}$. For the parameter $J_{x}\apprle2.5$,
the critical temperature has a minimum indicating the vestiges of
transition between $\mathrm{QFI^{-}}$ and $\mathrm{QFI^{+}}$, the
minimum disappears for $J_{x}>2.5$.

\begin{figure}
\includegraphics[scale=0.55]{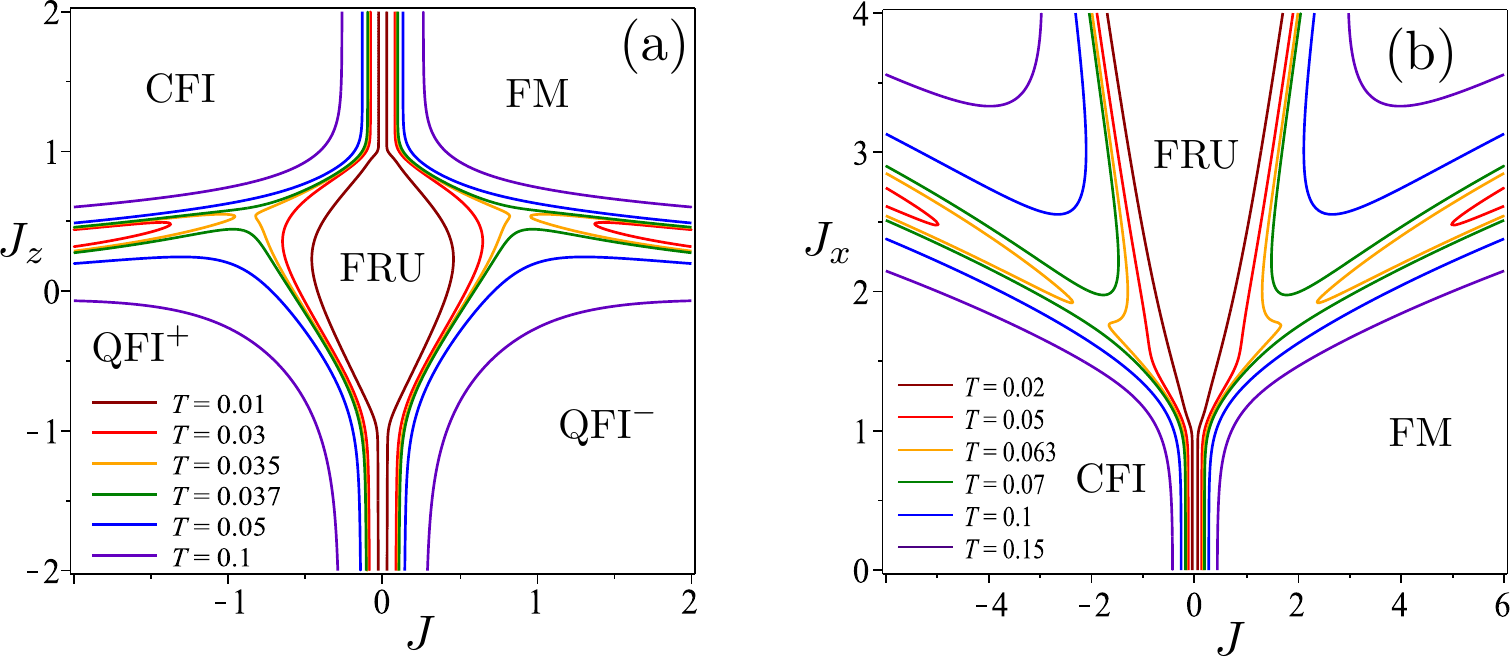}\caption{\label{fig:PHD-FT}(a) Isothermal curves in the plane $J-J_{z}$,
for fixed $J_{x}=1$ and a number of fixed temperature $T$ as denoted
inside. (b) Isothermal curves in the plane $J-J_{x}$, for fixed $J_{z}=1$
and several fixed temperature $T$ reported inside the panel.}
\end{figure}

In Fig.\ref{fig:PHD-FT}a is reported the isothermal curves in the
plane $J-J_{z}$, keeping fixed $J_{x}=1$. Here, we can identify
the influence of the zero-temperature phase diagram for several temperature
values. For detail, compare with Fig.\ref{fig: 0T phD}b. Likewise,
Fig.\ref{fig:PHD-FT}b displays the isothermal curves in the plane
$J-J_{x}$, considering fixed $J_{z}=1$, and for a set of temperatures
reported inside the panel.

\subsection{Spontaneous magnetization }

Another important quantity to be analyzed herein is the magnetic property
of the present model. That is why we need to examine the magnetization
of the decorated Ising-XXZ lattice. The Ising spin magnetization $m^{I}=\langle s\rangle$,
could be determined according to the magnetization of the effective
Ising model on a 2D triangular lattice. In contrast, the magnetization
of Heisenberg spins can be provided using the generalized star-triangle
transformation approach\citep{Fisher,syozi,PhyscA-09,Roj-sou11}
\begin{equation}
\langle\sigma_{k}^{z}\rangle=\eta_{k}\langle s_{1}\rangle+\frac{\gamma_{k}}{3}\langle s_{1}s_{2}s_{3}\rangle,\quad\text{with}\quad k=\{0,1\};
\end{equation}
which combines the single Ising spins thermal average $\langle s\rangle$
and triple Ising spins $\langle s_{1}s_{2}s_{3}\rangle$ linearly.
Hence we denote by $m_{0}^{H}=\langle\sigma_{0}^{z}\rangle$ the central
Heisenberg spin magnetization, and $m_{1}^{H}=\langle\sigma_{1}^{z}\rangle$
means the outer (decorated) Heisenberg spin magnetization. To find
the coefficients $\eta_{k}$ and $\gamma_{k}$, we use the following
relation\citep{Fisher,syozi,PhyscA-09,Roj-sou11}
\begin{equation}
\zeta(\varsigma,\sigma_{k}^{z})=\bigl[\eta_{k}(s_{1}+s_{2}+s_{3})+\gamma_{k}s_{1}s_{2}s_{3}\bigr]w(\{\boldsymbol{\sigma},s\}).\label{eq:Zeta-o}
\end{equation}
Analogously, for the Ising-XXZ model on a decorated honeycomb lattice
might be expressed as
\begin{equation}
\tilde{\zeta}(\sigma_{k}^{z})=\mathrm{tr}_{\{\sigma\}}\left[\hat{\sigma}_{k}^{z}\mathbf{V}(\{\boldsymbol{\sigma},s\})\right].
\end{equation}

However, the Ising spin coupling configurations $\{+++\}$ and $\{++-\}$,
are denoted merely as $\tilde{\zeta}(1/2,\sigma_{k}^{z})=\tilde{\zeta}_{1}(\sigma_{k}^{z})$
and $\zeta(3/2,\sigma_{k}^{z})=\zeta_{3}(\sigma_{k}^{z})$. Therefore,
these coefficients should be expressed as follow
\begin{alignat}{1}
\tilde{\zeta}_{3}(\sigma_{k}^{z})= & \mathrm{tr}_{\{\sigma\}}\left[\mathbf{U}_{3}\hat{\sigma}_{k}^{z}\mathbf{U}_{3}^{-1}\mathbf{\boldsymbol{\lambda}}_{3}\right],\label{eq:zeta3}\\
\tilde{\zeta}_{1}(\sigma_{k}^{z})= & \mathrm{tr}_{\{\sigma\}}\left[\mathbf{U}_{1}\hat{\sigma}_{k}^{z}\mathbf{U}_{1}^{-1}\mathbf{\boldsymbol{\lambda}}_{1}\right],\label{eq:zeta1}
\end{alignat}
where the orthogonal matrices $\mathbf{U}_{3}$ and $\mathbf{U}_{1}$
are obtained directly from the corresponding eigenvectors of the eigenvalues
provided in Table \ref{tab:Heisenberg-spin-eigenvalues,}, for each
configuration $\{+++\}$ and $\{++-\}$, respectively. Although, here,
we do not explicitly write the matrices $\mathbf{U}_{3}$ and $\mathbf{U}_{1}$,
because their dimensions are $16\times16$, and most elements of the
matrices are huge expressions. Thus, it is unnecessary to write down
here explicitly the matrices $\mathbf{U}_{3}$ and $\mathbf{U}_{1}$,
but we can easily generate them in algebraic programs.

From Eq. \eqref{eq:Zeta-o} we also have the following relations
\begin{alignat}{1}
\zeta_{3}(\sigma_{k}^{z})= & \left(\frac{3}{2}\eta_{k}+\frac{1}{8}\gamma_{k}\right)w_{3},\\
\zeta_{1}(\sigma_{k}^{z})= & \left(\frac{1}{2}\eta_{k}-\frac{1}{8}\gamma_{k}\right)w_{1}.
\end{alignat}
Hence the unknown coefficients results in
\begin{alignat}{1}
\eta_{k}= & \frac{1}{2}\left(\frac{\zeta_{3}(\sigma_{k}^{z})}{w_{3}}+\frac{\zeta_{1}(\sigma_{k}^{z})}{w_{1}}\right),\\
\gamma_{k}= & 2\left(\frac{\zeta_{3}(\sigma_{k}^{z})}{w_{3}}-3\frac{\zeta_{1}(\sigma_{k}^{z})}{w_{1}}\right),
\end{alignat}
with $\zeta_{3}(\sigma_{k}^{z})=\tilde{\zeta}_{3}(\sigma_{k}^{z})$
and $\zeta_{1}(\sigma_{k}^{z})=\tilde{\zeta}_{1}(\sigma_{k}^{z})$.
Here $\tilde{\zeta}_{3}(\sigma_{k}^{z})$ and $\tilde{\zeta}_{1}(\sigma_{k}^{z})$
are given by Eqs.\eqref{eq:zeta3} and \eqref{eq:zeta1}, respectively.

Furthermore, the thermal average relationship of single and triple
Ising spins was found in Ref.\citep{barry} which is rewritten as
follows:
\begin{equation}
\langle s_{1}s_{2}s_{3}\rangle=G(r)\langle s_{1}\rangle,
\end{equation}
where $G(r)$ after some algebraic manipulation becomes
\begin{equation}
G(r)=\frac{3r^{2}-3-2r\sqrt{\left(r-1\right)\left(r+3\right)}}{4(r-1)^{2}}.
\end{equation}
Moreover, the thermal average of Ising spin $\langle s_{1}\rangle$\citep{barry}
is given by
\begin{equation}
\langle s_{1}\rangle=\begin{cases}
\frac{1}{2}\left[\frac{\left(r-3\right)\left(r+1\right)^{3}}{\left(r+3\right)\left(r-1\right)^{3}}\right]^{\frac{1}{8}}, & r>3\\
0, & 0<r\leqslant3
\end{cases}.\label{eq:s0}
\end{equation}

Therefore, the thermal average of Heisenberg spins for both expressions
can be derived by using the following relation
\begin{alignat}{1}
\langle\sigma_{k}^{z}\rangle= & \left\{ \!\left(\tfrac{3}{2}+2G\right)\frac{\zeta_{3}(\sigma_{k}^{z})}{w_{3}}\!+\!\left(\tfrac{3}{2}-6G\right)\frac{\zeta_{1}(\sigma_{k}^{z})}{w_{1}}\right\} \!\langle s_{1}\rangle.\label{eq:M_h-t}
\end{alignat}

Note that the magnetization exponent satisfies the same universality
class of the two-dimensional Ising model, because all Ising ferromagnets
in two-dimensional have the same critical exponent of 1/8, regardless
of lattice.

The zero-temperature Heisenberg spin magnetization for the $\mathrm{QFI}^{\pm}$
regions occurs when $r\rightarrow\infty$, which means $G(\infty)=1/4$.
Under this condition simplifying \eqref{eq:M_h-t}, we have
\begin{equation}
m_{k}^{H}\!=\!\lim_{r\rightarrow\infty}\!\frac{\zeta_{3}(\sigma_{k}^{z})}{w_{3}}\langle s_{1}\rangle\!=\!\begin{cases}
\!-\frac{J\pm2J_{z}}{\sqrt{A_{\pm}}}\langle s_{1}\rangle, & k\!=\!0\\
\!\left(\!\frac{J\pm2J_{z}}{3\sqrt{A_{\pm}}}\mp\frac{2}{3}\!\right)\!\langle s_{1}\rangle,\! & k\!=\!1
\end{cases}\label{eq:Lim-m-H-FI}
\end{equation}
where $\pm$ corresponds to $\mathrm{QFI}^{\pm}$, respectively, and
$\langle s_{1}\rangle=\frac{1}{2}$.

\begin{figure}
\includegraphics[scale=0.54]{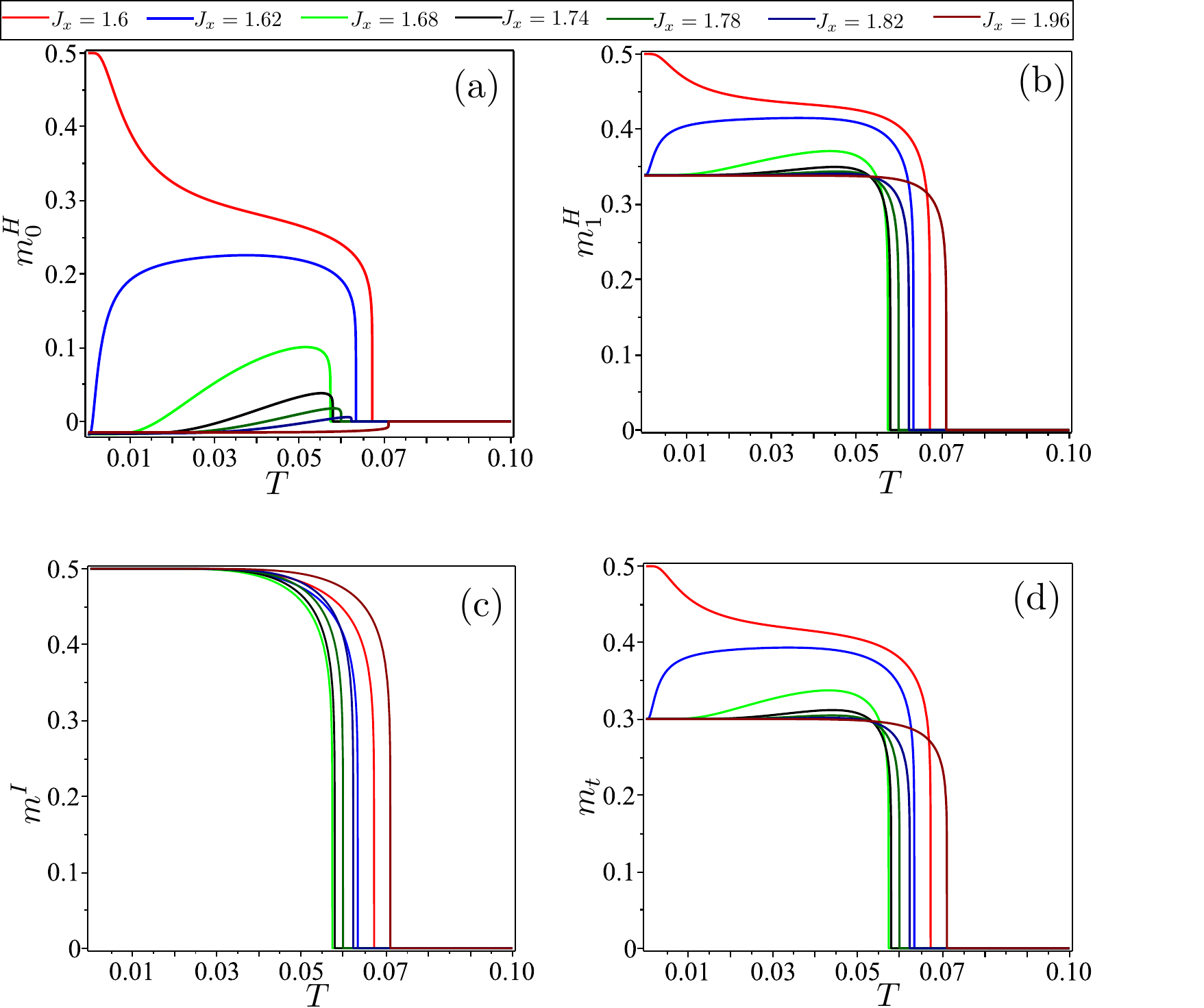}\caption{\label{fig:Magn-T}(a) Heisenberg spin magnetization $m_{0}^{H}$
as a function of temperature assuming fixed $J=2$, $J_{z}=0.9$,
and for several values of $J_{x}$. (b) Heisenberg spin magnetization
$m_{1}^{H}$ as a function of temperature for the same set of parameters
in panel (a). (c) Ising spin magnetization $m^{I}$ as a function
of temperature, assuming the same set of parameters considered in
panel (a). (d) Total spin magnetization $m_{t}$ as a function of
temperature for the parameters considered in panel (a).}
\end{figure}

In Fig.\ref{fig:Magn-T}a is depicted the central Heisenberg spin
magnetization $m_{0}^{H}$ as a function of temperature, assuming
fixed $J=2$, $J_{z}=0.9$, and for several values of $J_{x}$, as
described at the top of the panels. The zero-temperature magnetization
for central Heisenberg spins is given by Eq.\eqref{eq:Lim-m-H-FI}
for $J=2$, $J_{z}=0.9$ the curves leads to $m_{0}^{H}=-\frac{1}{2}\sqrt{\frac{3}{4J_{x}^{2}+3}}$.
Whereas in panel (b) is illustrated the Heisenberg spin magnetization
$m_{1}^{H}$ which bonds with Ising spin as a function of temperature.
The zero temperature magnetization is obtained from Eq.\eqref{eq:Lim-m-H-FI}
for $J=2$, $J_{z}=0.9$ leading to $m_{1}^{H}=\left(\frac{1}{6}\sqrt{\frac{3}{4J_{x}^{2}+3}}+\frac{1}{3}\right)$.
In panel (c) is depicted the Ising spin magnetization $m^{I}=\langle s_{1}\rangle$
as a function of temperature. So basically, observe a typical spontaneous
magnetization of the 2D Ising model. Finally, In panel (d), we observe
the total spin magnetization $m_{t}$ as a function of temperature.

\begin{figure}
\includegraphics[scale=0.46]{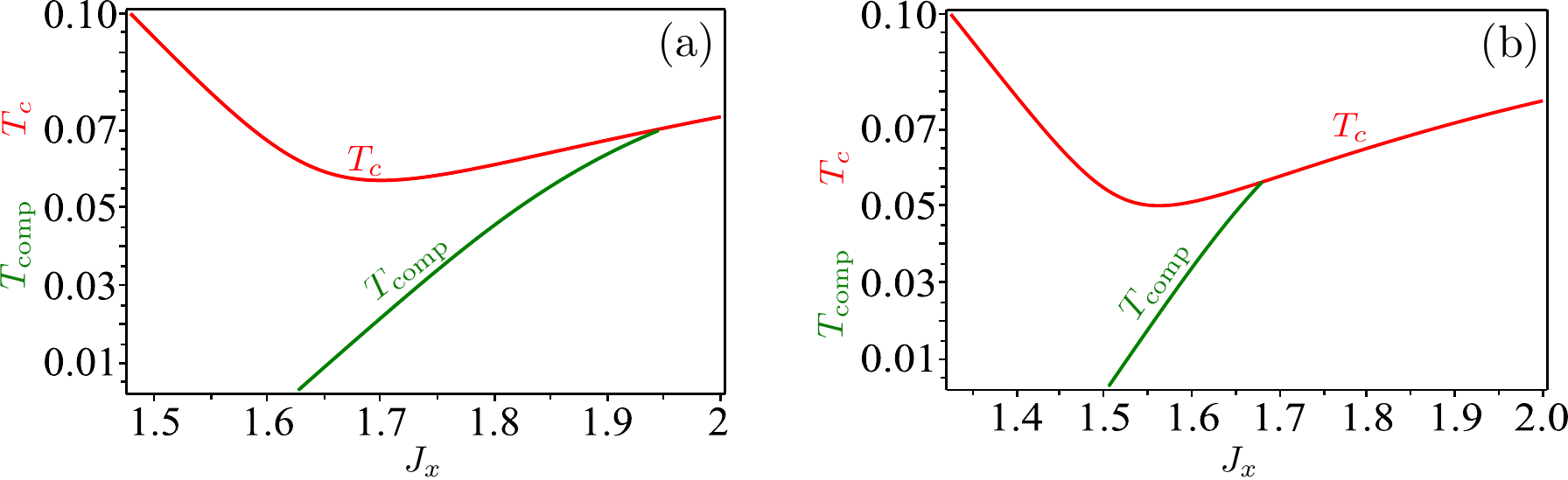}\caption{\label{fig:Comp} Compensation temperature for central Heisenberg
spin magnetization $m_{0}^{H}$ as a function of $J_{x}$, and fixed
$J=2$. (a) For $J_{z}=0.9$, and in panel (b) for $J_{z}=0.8$. }
\end{figure}

Another remarkable feature to note is the magnetization $m_{0}^{H}$
for the compensation temperature $T_{\mathrm{comp}}$.

Such anomalous behavior can be obtained by considering that Eq.\eqref{eq:M_h-t}
is null and setting $r>3$. For simplicity, let us define $t_{k}=\frac{\zeta_{3}(\sigma_{k}^{z})}{\zeta_{1}(\sigma_{k}^{z})}$,
and writing in terms of $r$ and $t_{k}$. We have
\begin{equation}
t_{k}=\frac{\frac{3}{2}-6G}{\frac{3}{2}+2G}r.
\end{equation}
By simplifying the above relation, we get the following identity
\begin{equation}
3r^{2}t_{k}+2rt_{k}^{2}-3t_{k}^{2}-9t_{k}-9=0.
\end{equation}
In particular, this requirement is only fulfilled for $k=0$, as shown
in Fig.\ref{fig:Magn-T}a. 

The compensation temperature is also explored in Fig.\ref{fig:Comp}
as a function of $J_{x}$. Panel (a) depicts for fixed values $J=2$,
$J_{z}=0.9$, and panel (b) illustrates assuming $J=2$, $J_{z}=0.8$.
The central spin magnetization $m_{0}^{H}$ has a peculiar behavior,
where we observe a compensation temperature $T_{\mathrm{comp}}$ when
the magnetization $m_{0}^{H}$ becomes null. In both cases of Fig.\ref{fig:Comp},
the compensation temperature of $m_{0}^{H}$ vanishes as soon as the
critical temperature is reached.

\section{Conclusion}

The two-dimensional spin-1/2 Ising-Heisenberg model in a decorated
honeycomb lattice with Ising and Heisenberg type exchange interaction
constitutes alternating clusters in the decorated honeycomb lattice.
We have verified that we can solve this model exactly through a generalized
star-triangle transformation. The relevance of this model is its close
relationship to the fully spin-1/2 Heisenberg model on a honeycomb
lattice, since four of five particles in each unit cell are of the
Heisenberg spins. Initially, we study the zero-temperature phase diagram
and found a typical ferromagnetic phase ($\mathrm{FM}$), one classical
ferrimagnetic phase $\mathrm{CFI}$, two types of quantum ferrimagnetic
phases ($\mathrm{QFI}^{+}$ and $\mathrm{QFI}^{-}$) and one quantum
spin frustrated phase ($\mathrm{QFR}$). It is worth noting that the
$\mathrm{QFR}$ states originates exclusively due to $J_{x}$ exchange
interaction, which corresponds to non-fitting antiferromagnetic Ising
model in effective triangular lattice. 

Additionally, we explored certain relevant quantities such as the
zero-temperature phase diagram, thermodynamics, spontaneous magnetization,
and critical temperature under several conditions of the effective
spin-1/2 Ising model on a triangular lattice.

Most of our investigation is focused on the quantum spin frustrated
region. The Heisenberg exchange interaction strongly influences the
critical temperature, and this occurs mainly in the low-temperature
region. We also obtain the residual entropy for the quantum spin frustrated
region by taking zero-temperature, which coincides with the residual
entropy found by Wannier\cite{wannier} for the Ising model on the
triangular lattice. Similarly, we derive the zero-temperature magnetization
due to the existence of the quantum exchange interaction.

This work was partially supported by Brazilian agencies FAPEMIG and
CNPq.

\end{document}